# The Classical Theory of Supply and Demand

Sabiou M. Inoua[1] and Vernon L. Smith[2]

Chapman University

Abstract. This paper introduces and formalizes the classical view on supply and demand, which, we argue, has an integrity independent and distinct from the neoclassical theory. Demand and supply, before the marginal revolution, are defined not by an unobservable criterion such as a utility function, but by an observable monetary variable, the reservation price: the buyer's (maximum) willingness to pay (WTP) value (a potential price) and the seller's (minimum) willingness to accept (WTA) value (a potential price) at the marketplace. Market demand and supply are the cumulative distribution of the buyers' and sellers' reservation prices, respectively. This WTP-WTA classical view of supply and demand formed the means whereby market participants were motivated in experimental economics although experimentalists (trained in neoclassical economics) were not cognizant of their link to the past. On this foundation was erected a vast literature on the rules of trading for a host of institutions, modern and ancient. This paper documents textually this reappraisal of classical economics and then formalizes it mathematically. A follow-up paper will articulate a theory of market price formation rooted in this classical view on supply and demand and in experimental findings on market behavior.

## 1 Introduction

This paper introduces and formalizes the classical view on supply and demand, which, we argue, has an integrity independent and distinct from the neoclassical theory. The new school, as is well-known, replaced the old in the 1870s through a widespread acceptance of constrained utility maximization as a core principle of economics.[1] Yet a century later, utility maximization is proven to have no interesting implication for aggregate (market) demand behavior, not even the law of demand except under artificial, restrictive conditions. While this major aggregation problem of standard price theory (the 'anything goes' or SMD theorem[2]) is often simply if unintentionally evaded in most applied models through the representative-consumer simplification (Hildenbrand, 1983; Grandmont, 1987), or by falling back on some additive utility structure or other forms of cardinality (Arrow, 1986; A. Kirman, 1989; A. P. Kirman, 1992), a few mathematical economists explored a pathway out of it, which consists of investigating the law of market demand as a collective regularity holding by *integration* over the distribution of consumers' preferences or incomes, thus turning aggregation into the solution, rather than the problem, of the law of market demand: for example, income or wealth effects, which are the main issue in the arbitrariness of neoclassical market demand, can be shown to be well-behaved by aggregation over a diverse population of consumers (essentially by the law of large numbers).[3] Ironically, this is the way the law of demand was understood in classical economics; Marshall, for example, who—as further emphasized below (Section 2.1)—tried to revive the classical view on supply and demand, makes it clear that his clause of constant marginal-utility of wealth (oft-discussed but oft-misunderstood in modern commentaries) is inconsequential on the aggregate of many consumers, poor and rich combined (Marshall, [1890] 1920, pp. 15-16, 83). More important for our objective here, the old school articulated a price discovery process which found unexpected new meaning in experimental markets establishing their remarkable decentralized convergence properties; these properties were not and could not be predicted by neoclassical modelling (Chamberlin, 1948; V. L. Smith, 1962). The goal of this paper is to rehabilitate mathematically the classical view on supply and demand. The classical view is easy to understand by opposition to the neoclassical one that replaced it, but which introduced into economics a series of mischievous innovations. For example, it used to be taken for granted in economics that economic reality is discontinuous at the micro level: not only is economic decision (both demand and supply behavior) binary (to buy or not to buy, to sell or not to sell is the problem), but goods come in discrete units and their relations are binary; for example, a consumer choses between two substitutes, rather than substituting infinitesimal amounts of goods, which strictly speaking are empty concepts. (For example: is an infinitesimal amount of a diamond still a diamond? Or even water?) While many of the marginalists were well aware of this point through Cournot ([1838] 1897, p. 50), they nonetheless assume that economic variables are smooth so as to use the tools of differential calculus.

[1] Economic Science Institute, Chapman University, 1 University Drive, Orange, CA 92866, USA; inoua@chapman.edu

[2] Economic Science Institute, Chapman University, 1 University Drive, Orange, CA 92866, USA; vsmith@chapman.edu

Moreover, individual demand and supply, before the marginal revolution, are defined not by an unobservable criterion such as a utility function, but by an observable monetary variable, the reservation price: the buyer's (maximum) willingness to pay (WTP) value (a potential price) and the seller's (minimum) willingness to accept (WTA) value (a potential price) at the marketplace. The underlying concept in value theory, in other words, used to be, not pleasure or satisfaction in consuming a good (which is experienced, or not, after the fact of purchase), but the consumer's valuation of a good, the maximum the consumer would be willing to pay for the good given his expectation of the good's usefulness. Market demand and supply are simply the cumulative distribution of the buyers' and sellers' reservation prices, respectively. This WTP-WTA classical view of supply and demand formed the means whereby market participants were motivated in experimental economics although experimentalists (trained in neoclassical economics) were not cognizant of their link to the past. On this foundation was erected a vast literature on the rules of trading for a host of institutions, modern and ancient. (Holt, in Kagel &Roth, 1995, pp. 360-377.)

Classical economics is not commonly viewed in these terms; rather it is often reduced to a simplistic supply-side, cost or labor theory of value, that fails to explain even the basic water-diamond value paradox because it was thought to be lacking the concept of marginal utility (a misunderstanding and distraction; Inoua & Smith, 2020a). Therefore, we must briefly revisit the old literature to document our interpretation and to frame the analysis in Section 2, which derives the classical conception of supply and demand progressively and heuristically from the classical literature. The second part of this paper (Section 3-4) is a formal restatement of classical supply and demand. Section 3 formulates mathematically the supply side of classical economics and derives key propositions of classical value theory.[4] Section 4 derives the less known demand side, with special attention to the foundation of classical demand theory, as it is made explicit in the French classical literature following Adam Smith.

## 2 The classical methodology

Overall, the classical economists adopted a methodology that can be summarized in three principles:

*Principle 1: It is a realistic portrayal of a market economy based on astute observation of individual behaviors and interaction in the marketplace.*

*Principle 2: It derives from the acute observations and facts about the economy's deep emergent properties that are the collective unintended consequences of these latter, the results of human actions but not of human design.*[5]

*Principle 3: Supply and demand are classically given by an observable, operational, monetary value: the reservation price—the buyer's maximum willingness to pay (WTP) and the seller's minimum willingness to accept (WTA).*

This paper, which is part of a general rehabilitation of classical economics, deals more specifically with Principle 3, the classical conception of supply and demand.[6] It is thus situated within the authors' overall rehabilitation project: it emphasizes how supply and demand were viewed before the marginal revolution. Alfred Marshall attempted to reconcile this old view of supply and demand with the new-born marginalist school.

## 2.1 Marshall's revival of a key principle

Alfred Marshall's 'pairs of scissors' image is often invoked in an oversimplification of the history of modern economics divided into three phases: from the classical, supply-centered, cost or labor theory of value to the early neoclassical demand-centered, marginal-utility theory of value, and to Marshall's synthesis of these two one-sided views into a unified price theory, which, allegedly, became the foundation of contemporary economics. Yet Marshall actually holds a more subtle view of the history of economics: his 'pairs of scissors' metaphor was merely intended to put an end to an old, essentially metaphysical, controversy over the *ultimate cause of value*—a problem which consisted of deciding which one of the two, unanimously recognized,[7] basic causes of value, utility or cost, is the most primitive cause. Marshall's reading of the history of economics, at the time, is unique, in that he most clearly recognized what was really at stake during the marginal revolution. Though he accepted diminishing marginal utility (DMU) as central to value theory (making him a marginalist of course), yet he saw in Jevons's program a major setback from a core methodological principle of classical economics, which is often overlooked in modern commentaries. This principle consists of dealing, as regards individual economic decisions, not with the ultimate psychological forces driving them, but operationally with the *monetary sacrifices* that people make in order to satisfy them. Thus, the relevant concepts for demand theory, for example, are, not the ultimate psychological motivations behind demand decisions (desire, want, pleasure), which had defied any precise quantitative modeling, but the money prices consumers are willing to pay in order to acquire the desired goods. This most fundamental principle, is applied equally to the supply side, and to market price theory more generally. Investigated from the standpoint of people's feelings, the value attached to an object reflects ultimately the desire of possessing it and the effort in producing it (an object, in this sense, is valuable, the more it is desired, and the more difficult it is to produce, in terms of toil and trouble)[8]. But investigated from the standpoint of the monetary values (or prices) traders in a market are willing to pay in order to produce or consume a good, the market price simply balances the ordered set of higher values that buyers are willing to pay to possess the good, with the ordered set of lower values that sellers are willing to accept in order to produce it. This, as textually documented below (Section 2.2), was precisely how supply and demand were understood long before the marginal revolution.

Alfred Marshall, perceptively recognizing this classical methodology, credited its discovery to Adam Smith, whom he viewed as having launched an epoch in the history of economics when he built from this principle a value theory that unifies all of economics "by a clearer insight into the balancing and weighing, by means of money, of the desire for the possession of a thing on the one hand, and on the other of all the various efforts and self-denials which directly and indirectly contribute towards making it. Important as had been the steps that others had taken in this direction, the advance made by him was so great that he really opened out this new point of view, and by so doing made an epoch." ([1890] 1920, Appendix B, p. 759). It is in fact this principle for measuring motives that confers upon economics a special quantitative nature among the social sciences ([1890] 1920, Book I, Ch. II, p. 12). Thus, in reaction to the hedonistic marginal utilitarianism of Jevons and Walras, who make pleasure the fundamental motivating category of economics, Marshall reformulates it operationally—as did the classicists (and, as it was applied, unknowingly, in the first market experiments)—entirely in terms of WTP and WTA (or demand-price and supply-price, as Marshall calls them, because he wanted to relate that difference to the incentive for seller entry)[9]. This WTP-WTA approach to supply and demand frames value theory throughout the classical literature; we first emphasize the demand side, since it is the less known.

## 2.2 Classical demand and French contributions

The demand side of classical price theory is sketched in Adam Smith's *Lectures on Jurisprudence* ([1763] 1869), under 'Cheapness and Plenty', which prefigures the *Wealth of Nations* ([1776] 1904). In the magnum opus, he simply grants that the purpose and foundation of consumer demand is to satisfy need, and he expresses demand in terms of WTP. He then directly explains price formation from the competition (higgling and bargaining) among the sellers and buyers in a market (Ch. VII of Book I).

Adam Smith did not articulate demand theory in a systematic, explicit, and formal way; but this articulation, which will be made explicit later by his disciples, can be sketched simply. Utility, or the capacity of a good to satisfy a consumer's need, is classically treated, not in the abstract, but in the specific sense of use-value: the value that a person attaches to an object by virtue of this object's usefulness and measured by what the person is willing to pay to get the object. Demand being thus founded on use-value, and use-value being operationally WTP, Adam Smith, throughout *Wealth of Nations*, equates demand directly to 'those who are willing to pay' a price. Elaborating on his work, the other classical economists also recognized explicitly WTP as the relevant concept in classical demand theory. Malthus, for example, announced early in his *Principles* that 'demand will be represented and measured by the sacrifice in money which the demanders are willing and able to make in order to satisfy their wants.' (Malthus, [1820] 1836, p 62)

The systematic and formal presentation of this classical demand theory appears in the important yet often overlooked French contribution to classical value theory, notably the works of Germain Garnier, Jean-Baptiste Say, Augustin Cournot, and Jules Dupuit, but also the Italian Pellegrino Rossi, who succeeded J.-B. Say at College de France's economics chair.[10] We see and interpret the French classical literature as a substantive rather than a mere echo, or popularization, of British classical economics; nor do we see it as detached from the classical school and interpreted as an anticipation of marginal-utility theory. Cournot, firmly in the classical stream, can be said to have charted new directions, notably on supply theory, which prepared the ground for the transition to neoclassical economics (though Cournot's views are more nuanced than its interpretations)[11]. Jules Dupuit, refining an intuition of J.-B. Say, clarified that use-value corresponds more precisely, not to any WTP, but to maximum WTP reservation price. J.S. Mill reached the same conclusion, putting it more technically: 'Value in use […] is the ex-treme limit of value in exchange.' Or: 'the utility of a thing in the estimation of the purchaser, is the extreme limit of its exchange value [the maximum price the purchaser would be willing to pay].' (Mill [1848] 1909, bk. 3: chap. 1, sec. 2; chap. 2, sec. 1.) Although no classical economist stated it explicitly, a consumer's demand follows by definition of the consumer's valuation: willingness to buy any unit whose value is greater than the price. This basic inequality, as formally emphasized in Section 4, defines entirely the market demand function, which is simply the total number of units that are valued more than the price offered: namely the complementary distribution function of consumers' values.

As to the foundation of classical demand, namely the determinants of WTP itself, it is not utility in the absolute, but a mix of utility and wealth. Need, unlike wealth, is a primitive demand concept.[12] Consumers buy goods to satisfy a list of needs: to each need is associated a certain good (or collection of goods) that satisfies it. A consumer's demand decision is shaped by a pyramid or *hierarchy of needs*, a ranking of needs from the most urgent to the least urgent: broadly speaking, from necessities, conveniences, to luxuries and fancies.[13]

This hierarchy of needs is *the fundamental principle of classical demand theory*, as shown with great clarity by J.-B. Say ([1828] 2010, p. 368) and as recognized by Dupuit (1849, p. 15), quoting Rossi's exposition ([1840] 1865, Lesson 5, pp. 87-88). The hierarchy of needs is partly objective, even universal. Thus, no good is valued higher than water, which serves a vital biological need; a diamond, serving an ornamental need, is valued lower, since one would be willing to give all diamonds at hand to survive. Yet the market prices for a cup of water and a diamond (or their objective exchange values) are inversely related to their subjective valuations—a paradox which is long-known to be solved by the concept of *scarcity*, and which no classical economist regarded as an un-solved mystery (Inoua & Smith, 2020a). But the hierarchy of needs is in part subjective as well. At any rate, the economist takes a consumer's needs as given.

In summary, the classical value literature from Adam Smith to Jules Dupuit offers a consistent picture of demand, which Marshall attempted to reconcile with the newborn marginal school.[14]

## 2.3 Classical cost and supply decision

Cournot, who beautifully expounds the old view on demand in one fascinating paragraph ([1838] 1897, p. 50), goes on, however, to introduce an abstract theory of atomistic profit-maximizing firms—which will become the standard view on supply in the hands of the marginalists. Like utility maximization, profit maximization thus conceived raises decision problems that actual firms hardly face; for example, it leads to an ill-defined supply function when marginal cost is nonincreasing (most manufactured products?)[15] or zero (information products such as software programs). The basic problem lies in the definition of the firm's production possibility set as an unbounded set: thus, a firm facing a constant unit cost, for example, would be willing to supply an infinite amount of output when facing a price greater than the unit cost. Consider in contrast an actual firm under the same cost condition: granted that the firm's maximum production capacity is finite, its supply decision is obvious: willing to supply every producible unit at any price greater than the unit cost.

Prior to Cournot's innovation, the discussion on supply decision seems to presume none other than the basic principle, taken for granted, of willingness to supply whenever it is profitable to do so, that is, at any price beyond a minimum acceptable price (the minimum WTA). Like classical demand, classical supply is not explicitly and formally defined; yet it goes without saying, by definition of the concept of minimum price, which corresponds to the cost of production, or more precisely the cost of supplying the commodity to the market. Classical cost is the monetary evaluation of all the sacrifices the producer makes in order to supply the commodity, including the expectation of a minimum profit compensation (without which none would engage in the toil and trouble of producing). Regarding this *monetary cost valuation*, Adam Smith, and this is crucial, assumes none other than the actual practice of firms in their cost accounting: not only the obvious part, wages and the cost of raw materials (determined by market rates), but also the monetary estimate of the use of fixed capital (allowance for depreciation), and the producer's minimum compensation, which is a subjective evaluation in general; but would expect at least the 'ordinary profit' in the industry. Adam Smith, who particularly emphasizes this actual practice of firms on many occasions, faces none of the technical complications of later authors who will treat capital as a physical agent of production (the aggregation problem as it applies to fixed capital).[16] Granted the cost valuation (a monetary evaluation of all expenses of production), which is the producer's minimum WTA, the producer is willing to supply any unit at a price that covers at least the money cost: willingness to sell any unit whose cost is lower than the price. This inequality, as emphasized below, defines entirely the supply function: since market supply is the number of units that can be supplied profitably, it is given by the cumulative distribution function of the unit costs.

## 2.4 Limitations of Marshall's synthesis

Marshall's insightful attempt to revive the classical method does not pay full justice to the old paradigm, due to his desire to integrate neoclassical utility theory into his treatment. For example, he defines the demand function as the surplus-maximizing quantity of a representative consumer that hypothetically buy by infinitesimal increments of the commodity, and he founded the law of demand on diminishing marginal WTP (or 'demand price'); he defines supply similarly. This and other neoclassical imports play no essential role in the classical formulation, and none when Marshall is brought to a description of price formation in a local country market. [Marshall ([1890] 1920, Book V, Chap. II) simply reverts to a WTP/WTA description of "higgling and bargaining" as we find it in A. Smith ([1776] 1904, Book I, Ch. VII)).]

Marshall ([1890] 1920, p. 64) acknowledges that economic reality is discontinuous in regard not only to the quantity of goods but also of individual behavior, but demand is smoothed, "in so far as the motives of that action are measurable by a money price; and in these broad results the variety and the fickleness of individual action are merged in the comparatively regular aggregate of the action of many"; a form of the law of large numbers. The motivation for this theoretical procedure is this other equally important principle of classical methodology noted in the introduction, which Cournot expressed in a most fascinating way ([1838] 1897, p. 50), and which consists more generally of *investigating economic regularities as collective regularities emerging by aggregation over agents*. For the classical economists, this meant aggregation over the distribution of agents' characteristics, and not just a hypothetical average agent, as Marshall did.

Finally, we have Marshall's deviation from the old school in regards to his dealing with wealth effects through his oft-discussed clause of constant marginal utility of wealth, which may be erroneously interpreted to mean that thinking in terms of reservation price is a narrow case of the utility-function view; though Marshall in fact goes on to argue that variations in marginal utility of wealth, like the discontinuities of demand with respect to price variations, are of no significance on the aggregate of many consumers, with poor and rich, young and old combined (Marshall, [1890] 1920, pp. 15-16, 83). Space forbids to elaborate further on Marshall's synthesis. Rather this paper concerns the restoration of the old school, on its own merits, bereft of Marshall's attempted synthesis.

## 3 Supply

### 3.1 The market supply function

Consider a market economy in which $n$ goods and services (including labor services) are traded at market prices $\mathbf{p} = [p_1, ..., p_n]$. A producer in any market is willing to sell any unit that can be produced profitably. Consider a given commodity, which we single out by not indexing it. In terms of cost, each unit of a commodity produced is characterized by the number of inputs from other commodities its production required, which we denote generically as a vector[17] $\mathbf{a} = [a_1, ..., a_n]$, and the input prices. Thus, the cost of a unit of a commodity is

$$c = \sum_k a_k p_k = \mathbf{a}\mathbf{p}. \tag{1}$$

Different units clearly may cost differently to produce since they may involve a different mix $[\mathbf{a}, \mathbf{p}]$. The producer would be willing to sell any unit at a price $p \geq c$. The function that associates to any price vector the total number of units of that commodity that would be supplied, at this price vector, by all the producers in the market is, by definition, the market supply function. It is by construction the cumulative distribution function of unit costs, which is a non-decreasing step function of the market price, ceteris paribus:

$$S(\mathbf{a},\mathbf{p}) = \overline{S}F(\mathbf{a},\mathbf{p}), \tag{2}$$

where by definition $F(\mathbf{a},\mathbf{p}) = \mathbb{P}(c \leq p)$ and $\overline{S} = S(\mathbf{a},\infty)$ is the total (maximum) supply capacity in the market (the total number of units that can be supplied, which would be fully supplied were the market price infinite); for a particular market, the distribution in question refers to the collection of all units of the same commodity.

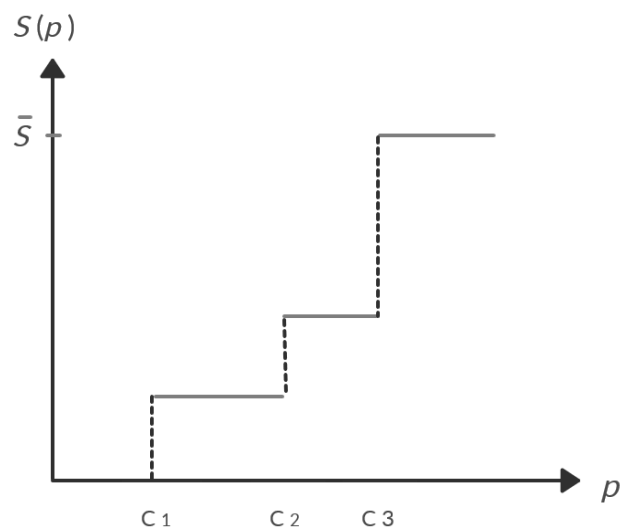

Figure 1: Market supply is the cumulative distribution of unit costs. The prices of inputs assumed fixed here: their changes corresponding to shifts of the supply curve.

The relevant unit of analysis in supply theory is the supply for a unit of a commodity: $S_j(p) = 1$ if $c_j \geq p,$ and $S_j(p) = 0,$ otherwise. The supply of a firm is simply is a list of such elementary supplies: it is entirely specified by the number of units the firm can produce and the corresponding list of costs for each unit.

### 3.2 The labor theory of value

Unlike Ricardo (and his followers), Adam Smith mentioned only passingly the labor theory of value which applies only in rude societies:

> "In that early and rude state of society which precedes both the
> accumulation of stock and the appropriation of land, the proportion
> between the quantities of labour necessary for acquiring different
> objects seems to be the only circumstance which can afford any rule for ex-
> changing them for one another. If among a nation of hunters…it usually costs
> twice the labour to kill a beaver which it does to kill a deer, one beaver should
> naturally exchange for or be worth two deer…the produce of two days or two
> hours labour, should be worth double…the produce of one day's or one hour's
> labour.[18] (Smith, [1976] 1904, p. 49)

Ricardo himself makes it clear that the labor theory of value applies only when goods can be produced with homogenous labor, in abundant amounts, and at proportional costs (which excludes the complications of the diversity of fixed capital that Ricardo would later deal with); Ricardo later concedes that quantity of labor is the dominant (not the sole determinant) of price, at least in the long run. Both J.-B. Say and Malthus oppose Ricardo even regarding long-run value, which, as they point out, is determined by both supply and demand. Although historically Ricardo is perhaps the most influential of the classical economists, it is a mistake to view Ricardo's formulation as a culmination of this school of thought; it is not true upon scrutiny that Ricardo was resolving inconsistencies in Adam Smith's view; Ricardo was dealing with technical complications (the diversity of labor and fixed capital) that are precisely the reason why Adam Smith mentions only passingly labor theory of value and goes on to develop (Book I, Ch. VII) the general theory of market price formation. For example, Ricardo deals with the diversity of labor following Adam Smith's view that wage differentials are determined by the 'higgling and bargaining' of the market; but this is precisely why the relevant price theory in general is a theory of supply and demand. While this point is now commonplace in economics, yet Adam Smith in most commentaries on classicalism is overshadowed by Ricardo; and the classical school is still commonly reduced to the labor theory of value, although much of the controversies that opposed Ricardo to both Say and Malthus pertains precisely to Ricardo's reduction of this school of thought and his downplaying the role of demand in price theory. Let it be reminded that the labor theory of value is equivalent to assuming a Leontief price system, as known since the influential revival of Ricardo's theory by Sraffa (1960); see Inoua and Smith (2020a).

## 4 Demand

### 4.1 The market demand function

For a given commodity, let $h_k = 1$ if a consumer considers commodity $k$ is more urgent than the commodity under consideration, and $h_k = 0$, otherwise; let $\mathbf{h} = [h_1, ..., h_n]$. Consumers differ in terms of hierarchy of needs $\mathbf{h}$ and wealth $w$. Let the distribution of these consumers' attributes be referred to as $[\mathbf{h}, w]$. A unit of a commodity will be demanded by a consumer if the money left out of his wealth, once more urgent needs are considered, can afford the unit: that is, if $(w - \sum_k h_k p_k) \geq p$. Thus the consumer's valuation of the commodity, as given by his maximum WTP for it (reservation price) is

$$v = w - \sum_k h_k p_k = w - \mathbf{h}\mathbf{p}.$$  (3)

The consumer would be willing to buy every unit whose $v \geq p$. The market demand function at any price is therefore the number of units of the good to which consumers attach a greater value than the price: it is by construction the complementary distribution function of the consumers' reservation prices, which is a non-increasing step function of price, ceteris paribus, which we write generically:

$$D(\mathbf{h}, w, \mathbf{p}) = \bar{D}G(\mathbf{h}, w, \mathbf{p}), \tag{4}$$

where $G(\mathbf{h}, w, \mathbf{p}) = \mathbb{P}(v \geq p)$ and $\bar{D} = D(\mathbf{h}, w, \mathbf{0})$ is the total (maximum) number of units of the good that consumers need (their overall demand were the market price zero).

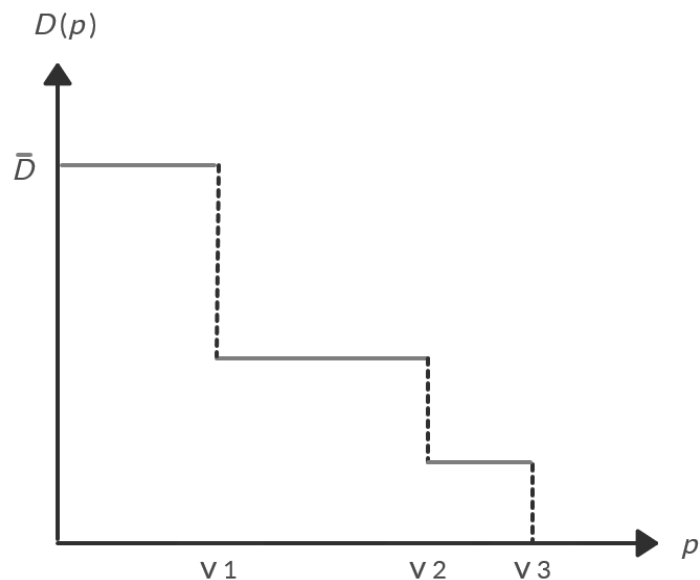

Figure 2: Market demand is the complementary distribution function of consumers' valuations (or reservation prices).

## 4.2 Remarks on a few conceptual distinctions

### *Quantity: needed versus demanded versus bought*

The focus on the elementary demand for a unit of a good irrespective of the identity of the demander, is on purpose, as unit demand is the relevant unit of analysis. The total demand of a consumer is merely a list of unit demands: it is entirely specified by the number of units the consumer needs and their reservation prices. Let $d_i = d_i(\mathbf{h}_i, W, \mathbf{p})$ be the consumer's demand function for commodity $i$. The demand (resp. supply) function, be it reminded, indicates the quantity the consumer would demand were the consumer facing any arbitrary price $\mathbf{p}$. The total quantity needed is $d_i = d_i(\mathbf{h}_i, W, \mathbf{0})$.[19] It is a different concept from the quantity the consumer succeeded to actually buy, which depends on the extent of competition in the market. The distinction more generally between quantity demanded or supplied (willingness to trade) and the quantity actually traded is of utmost importance: serious paradoxes and conceptual obstacles in the neoclassical theory of price formation can be shown to be due to failure to acknowledge explicitly actual trades as a conceptually district notion from a supply and demand function, which summarize all willingness to buy and sell at any possible price; since Walras, for example, it is common to treat supply and demand as if always fulfilled into actual trades, which would be the case only in equilibrium.

### *Complements and substitutes*

Realistically, the relationship among commodities are binary. Consider two commodities $i$ and $k$ viewed from the consumer's viewpoint. By definition, the two goods are complementary if they jointly serve the same need, and hence are jointly demanded: $d_i > 0$ implies $d_k > 0$, and vice versa. They are substitutes if they serve the same need, but interchangeably, so that demand for one excludes demand for the other: $d_i > 0$ implies $d_k = 0$, and vice versa. In both cases, demand for one good derives logically from demand for the other: theoretically, therefore, all substitutes, on the one hand, and all complements, on the other, can be treated mathematically as forming one class of goods, of which knowledge of one element is like knowledge of the whole. Thus, we are left with the hierarchical relation among goods as the essential notion in demand decision: good $i$ is more urgent than good $k$ if $d_i = 0$ implies $d_k = 0$.

### 4.3 The pyramidal model of market demand

The law of demand and the law of supply, as we saw (Figure 1 and Figure 2), hold by construction, in the sense that market demand and market supply are, respectively, nonincreasing and nondecreasing (step) functions of price, ceteris paribus. This weak version of the two laws is all that the theory of price formation requires in general (Inoua & Smith, 2020b). But stronger versions can also be derived under minimum assumptions, as the French classical[20] economists emphasize; we derive them formally now.

For partial-equilibrium purposes, let wealth distribution and the prices of related goods be given. In a sufficiently large market, as Cournot beautifully emphasized ([1838] 1897, p. 50), market demand (resp. market supply) can be assumed to be a smoothly decreasing (increasing) function of price. A large market can be formally defined as an idealized version of a market that involves a sufficiently large number of distinct values and costs modeled by continuous density functions supported on continuums. Formally, a large market is therefore one for which the distribution of costs and values are modeled by the continuous density functions $F'$ and $-G'$ supported on the intervals $[c_{\min}, c_{\max}]$ and $[v_{\min}, v_{\max}]$ respectively, over which the density functions are strictly positive by definition. It then follows by construction $\partial S/\partial p = \bar{S}F'(p) > 0$ and $\partial D/\partial p = \bar{D}G'(p) < 0$ on the respective supports.

Now, consider a market in isolation, setting $\mathbf{hp} = 0$ in (3), so that $v = w$ and $G(\mathbf{h}, w, \mathbf{p}) = G(p) = \mathbb{P}(w \geq p)$, so that market demand is given by the distribution of wealth across consumers, which Garnier, Say, and Dupuit represented as a pyramid (Figure 3), whose top represents the wealthy minority and whose base represents the poor majority: more generally, the pyramid represents the distribution, not of wealth per se, but of the portion of wealth each consumer would be willing to pay for the commodity (Garnier, [1796] 1846, pp. 195-196; Say, [1828] 2010, p. 370, footnote 1). The pyramidal assumption in formal and general terms simply means a decreasing probability density of wealth or WTP more generally, $-G'' < 0$, so that market demand is a convex function of price, as Dupuit insightfully observed (1844, pp. 367-368). Unlike the law of demand proper, however, this convexity property, sometimes referred to as Dupuit's second 'law' of demand (Ekelund Jr & Thornton, 1991; Humphrey, 1992 )[21], is true only to the extent that the pyramidal assumption is true: it is not essential to price theory.

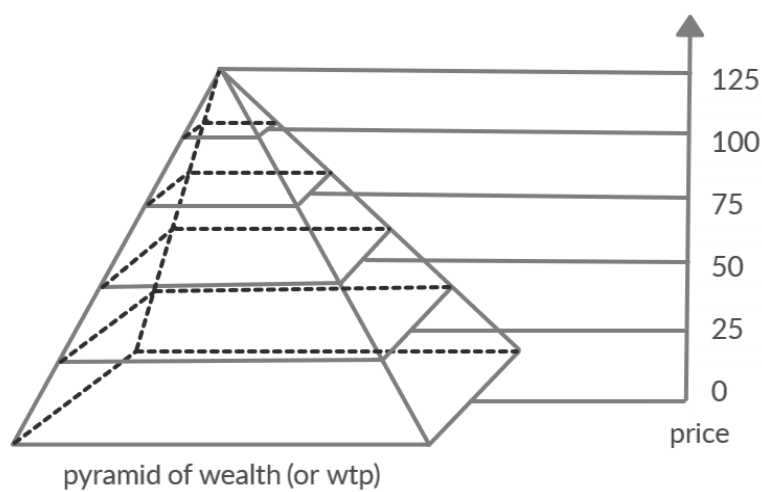

pyramid of wealth (or wtp)

Figure 3: Pyramidal model of market demand (Garnier, Say, Dupuit). The pyramid (left) represents the distribution in society of wealth (top=the wealthiest, bottom=the poorest), or more precisely the distribution of consumers' WTP. Market demand, at each price, is measured by the cross-sectional area of the pyramid corresponding to the price. If price is zero, all consumers can afford the good; as price increases, a lower and lower fraction of society can afford the good; and beyond some maximum value (125), none can afford the good.

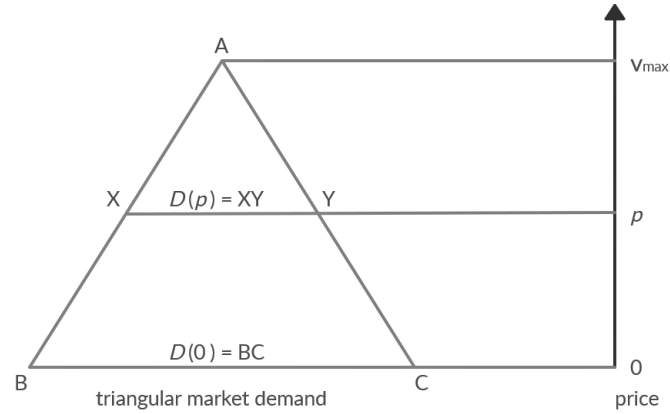

Figure 4: Triangular market demand model (2-dimensional version the pyramidal model): market demand is the length of the segment XY: it is a linear function (as can be proven from elementary geometry). The pyramidal market demand (3-dimentiomal) is simply the square of this triangular market demand.

For illustration, assume (following Garnier, Say, and Dupuit) a square pyramid (Figure 3): then market demand is simply the square of its two-dimensional image (Figure 4), obtained by reducing the pyramid to a triangle, and the cross-sectional areas of the pyramid to segment lengths. It can be shown from elementary geometry [Thales's (intercept) and Pythagoras's theorems] that

$$G(p) \equiv \frac{D(p)}{D(0)} = \frac{XY}{BC} = \frac{v_{\max} - p}{v_{\max}}. \tag{5}$$

This corresponds to a uniform probability distribution of consumers' reservation prices supported on $[0, v_{\max}]$. [The extension to the general case $[v_{\min}, v_{\max}]$ is straightforward: $G(p) = (v_{\max} - p)/(v_{\max} - v_{\min})$.] The two-dimensional, pyramidal, model yields (assuming again a square pyramid)

$$G(p) = (\frac{v_{\max} - p}{v_{\max}})^2, \tag{6}$$

which is indeed a (strictly) convex demand function. It should be insisted that the law of demand, even in its smooth version, holds for any continuous probability distribution of consumer valuations; it is only the secondary, convexity, property that requires the pyramidal assumption $G'' > 0$.[22]

Although this probabilistic view on demand (based on the distribution of consumers, as ranked in different orders of society based on WTP, or, as we would say today, in different statistical classes), was first formalized in the French literature, it was in fact implicit throughout the classical school: thus Adam Smith was implicitly treating consumers in that way when, for example, he observed in his *Lectures* that 'everything is dearer or cheaper according as it is the purchase of a higher or lower set of people' (A. Smith, [1763] 1869, p. 177).

**5 Conclusion**

The classical economists mostly adopted the same realistic view of the market economy; not that they concurred to it by an explicit a priori methodological commitment; but, rather, they concurred to it because they were essentially adopting the same mentalizing process, which consists, first of all, of carefully observing everyday economic life, and then deriving from this acute observation, deep emergent regularities that are unintended consequences of these ordinary individual behaviors and interactions. J.-B. Say explained this classical methodology with great clarity in his *Cours complet* (and complained that Ricardo at times deviated from it).[23]

But in place of this realistic methodology, the early neoclassical economists substituted a-priori axiomatic theorizing, whereby a theorist starts beforehand with a set of axioms (for example: pleasure explains every move in human behavior) and from these abstract premises, constructs through a chain of formal deductions an imaginary economy (often populated by a single player: Robinson Crusoe); and even after the axioms have proved, upon more scrutiny, to be empirically empty by the barrenness of their implications, the as-if theorist finds further refuge in fiction, and an excuse for holding onto these axioms. Thus, price-taking behavior, for example, was recognized from the very beginning to be a dead end as a premise for a theory of competitive market price formation (for if everyone in the economy takes price as given, where do these prices come from in the first place?) Yet in the face of this dead end, an early evasion simply assumed a perfect market in which supply and demand and the consequent equilibrium price are perfect knowledge to every trader beforehand (Jevons, [1871] 1888). A second, now-standard, escape consists of simply postulating the existence of a fictional auctioneer who seeks all equilibrium prices by trial and error (Walras, [1874] 1954).[24] In the same spirit, the aggregation problem of neoclassical demand is evaded through the representative-consumer assumption.

---

[1] In the standard undergraduate intermediate theory course the consumer chooses units (x1, x2) of goods defined on a continuous commodity space to Max U (x1, x2) [increasing and concave in (x1, x2)] subject to I = p1x1 + p2x2, given (U, I, p1, p2). Every aspect of this model is contrary to the classical economic model, wherein: (1) the consumer chooses only discrete units of goods [the primary meaningful application to continuous action spaces is in finance]; (2) if income is fixed and constrains choice, wealth is stationery, but the classical economists saw as their end and purpose to inquire as to the nature and causes of the wealth of nations; (3) prices and income were to be determined in the market, and were not given to it; (4*) U was* a hidden variable to the classics, but people in markets revealed that

they had willingness to pay demand valuations for goods, and for inputs to supply goods, and these were central to their analysis.

[2] That is, the Sonnenschein-Mantel-Debreu theorem (Sonnenschein, 1972, 1973a, 1973b; Debreu, 1974; Mantel, 1974; Shafer & Sonnenschein, 1993; Rizvi, 2006).

[3] This abstract revival of the law of demand as an aggregate regularity, systematically explored in Hildebrand (1994), can be viewed as part of the general 'regularity by aggregation' literature, which seeks to solve the multiple indeterminacy of neoclassical theory (e.g. the indeterminacy of demand when preferences are non-convex). For an overview of this literature, see Trockel (1984).

The probabilistic view on demand has in fact resurfaced in different other forms even before the SMD crisis; for example, Becker's intuition that even impulsive or random consumer choice constrained by a budget would obey the law of demand by aggregation, independently of utility maximization (Becker, 1962); for a recent revival of this view, see (Shaikh, 2012); for a recent experimental exploration in a general-equilibrium context, see Crockett, Friedman, and Oprea (2019).

[4] Classical market price theory is not to be confounded with the labor theory of value, on which Ricardo insisted, and which is a diversion from the general classical theory of price formation; the precise conditions under which the labor value theory applies are restated in 3.2; for a formal derivation see Inoua and Smith (2020a).

[5] "[N]ations stumble upon establishments, which are indeed the result of human action, but not the execution of any human design" (Ferguson, 1782, p 205)

[6] A follow-up paper will present a theory of price formation that is rooted in the classical view on competition and relates that view to experimental findings on market behavior. Other more preliminary papers tackle obstacles and limitations of the classical literature, which may seriously impede or even discourage the modern reader's assessment of the old literature, and explains the articulation of value theory in the classical school: the technical jargon of classical economics (natural price, monopoly price, effectual

demand); the endless classical controversies regarding essentially unsolvable, metaphysical, issues (the invariable measure of value and the ultimate cause of value); these controversies can mislead the modern reader into seeing irreconcilable divergences in the classical school (and whose unity may thereby be questioned).

[7] None of the protagonists of this old utility-versus-cost controversy denied that market price is determined jointly by utility and cost: 'Almost all writers have agreed substantially, and have rightly agreed, in founding exchangeable value upon two elements, -power in the article valued to meet some natural desire or some casual purpose of man [utility], in the first place, and, in the second place, upon difficulty of attainment [cost]. These two elements must meet, must come into combination, before any value in exchange can be established.' (De Quincey, 1844, p. 13).

[8] Following the classical terminology, Marshall ([1890] 1920, p. 282) distinguishes between real cost (pain, effort, difficulty of producing a product) from money cost (the monetary valuation of the real cost: the expenses of production, including a minimum profit requirement). Today we take for granted the fact that cost of production (difficulty of production) is measured by the monetary sacrifices the producer makes (the expense of production). Yet this principle (which applies to the demand side as well) will be adopted in the neoclassical school only as a shortcut or a concession, Thus, Jevons's program, a pure subjectivism, aimed at explaining value entirely in terms of pleasure (utility, demand) and its negation, pain or effort (disutility of labor, supply).

[9] Marshall explains this often-overlooked classical principle in a systematic way throughout his famous *Principles of Economics* ([1890] 1920), particularly in Book I, Ch. II, and makes it clear that its paternity originates with Adam Smith, and his "unsurpassed powers of observation". Marshall did not realize, however, that all the classical economists, as they followed in Smith's footsteps, reached the same conclusion, that value theory should be founded on people's monetary valuations. With hindsight, it was a major editorial mistake on the part of Marshall to have moved the section on the

history of economics—in which he clearly explains the classical paternity of this principle—to the Appendix, in response to the public demand for making the first parts of his book less tedious.

[10] The references are G. Garnier ([1796] 1846, pp. 195-196); J.-B. Say (1803 [2006], vol. 2, bk. 2, chap. 1; [1828] 2010, vol. 1, part 3, chap. 4), particularly the later book, the *Cours complet*, which synthetizes and extends the material covered in Say's earlier books; P. Rossi ([1840] 1865, Vol. I, Lesson 5); A. A. Cournot ([1838] 1897, chap. 4), and J. Dupuit (1844, 1849). Many of the relevant passages of this French literature on demand and value are quoted in (Ekelund Jr & Hébert, 1999).

[11] On more of the methodological innovations of Cournot, see Smith and Inoua (2019).

[12] The classical concept of WTP is out of wealth, not income. The idea of income as a constraint on commodity choice is conceptually a blatant neoclassical error. Modeling choice in the current period only makes static sense if one of the goods is variable and constitutes saving—not consuming—with personal value in the current period; otherwise, the action set is not closed and part of the dynamics of wealth accumulation. For Adam Smith that value is security, which he saw as protection against downside loss: "We suffer more…when we fall from a better to a worse situation, than we ever enjoy when we rise from a worse to a better. Security, therefore, is the first and the principal object of prudence." (Smith, 1759, p. 213)

[13] More than a century later, A. Maslow (1943) offers a famous psychological theory of the pyramid of needs.

[14] Marshall perhaps first noticed the classical methodology through his reading of Cournot and Dupuit, whose influences on him he acknowledged ([1890] 1920, p. 85, footnote 1). The remarkable thing is his seeing the connection with the classical school more generally and tracing its origin back to Adam Smith.

[15] It was common belief in the classical literature that most (manufactured) goods are produced at constant or decreasing unit costs (production on a

large scale leads to efficiency gains because it promotes a better division of labor, for example); only agricultural produce and mined resources were believed to command increasing unit costs. See, for example, Mill ([1848] 1965, p. 464 ff.) Various modern surveys suggests that increasing marginal costs are indeed exceptional in practice: it seems, according to one survey of the US economy, that 'only 11 percent of GDP is produced under conditions of rising marginal cost.' (Blinder, Canetti, Lebow, & Rudd, 1998) For a discussion of these issues with neoclassical supply theory, see Keen (2011, Part I, Ch. 5).

[16] This assessment of the producer's minimum price through a realistic cost valuation is clearly stated throughout *Wealth of Nations*, starting from Ch. 7 of Book I ([1776] 1904, p. 50). It is repeated countless times in specific contexts. For example: 'When any expensive machine is erected the extraordinary work to be performed by it before it is worn out, it must be expected, will replace the capital laid out upon it, with at least the ordinary profits.' (p. 103) Or elsewhere: 'The lowest price at which coals can be sold for any considerable time, is, like that of all other commodities, the price which is barely sufficient to replace, together with its ordinary profits, the stock which must be employed in bringing them to market.' (p. 168) This view would appear throughout the classical literature. Marshall ( [1890] 1920, p. 299) also clarifies this cost valuation of firms.

[17] More precisely a matrix with double entry, in more explicit notation; but by fixing the commodity under study, the notation is simplified from distracting indices.

[18] A source of confusion, perhaps, is the distinction between the labour theory of value and the concept of labour as a measure of value, developed by Smith. Thus, "The value of any commodity…to the person who possesses it, and who means not to use or consume it himself, but to exchange it for other commodities, is equal to the quantity of labour which it enables him to purchase or command. Labour, therefore, is the real measure of the exchangeable value of all commodities." (Smith [1776, 1904, p 32)

[19] Adam Smith introduces the distinction 'absolute demand' versus 'effectual demand' to reflect the nuance between quantity needed or desired and quantity effectively demanded (determined by need, constrained by wealth). An explanation of the technical classical jargon (absolute versus effectual demand; natural versus monopoly price, etc.) can be found in Inoua and Smith (2020a).

[20] By now, there should be no reason for hesitating to refer to all of them simply as classical economists.

[21] Dupuit emphasized, besides the standard law of demand proper (as price drops, quantity demanded increases), a so-called second 'law' (Ekelund Jr & Thornton, 1991; Humphrey, 1992 ): the increase in demand due to a price drop is the higher, the lower the initial price: that is, the second derivative of the demand function is positive, which, as Dupuit justified intuitively, and as formally proven in the text, derives from the pyramidal assumption.

[22] This nuance is missing in the original literature, and understandably so, since back then even the basic difference between continuity and differentiability (let alone probability density function versus cumulative probability function) have yet to be well-understood, as Cournot's characterization of continuity at one point in his book attests ([1838] 1897, p. 50, phrase italicized).

[23] J.-B. Say discusses the methodology underlying classical economics in the opening *Considerations générales* of the *Cours complet* ([1828] 2010, pp. 3-61).

[24] Walras merely set the stage for this fiction: the explicit introduction of the imaginary auctioneer in the theory of tatonnement came later.